\documentclass[sigconf]{acmart}
\AtBeginDocument{%
  \providecommand\BibTeX{{%
    \normalfont B\kern-0.5em{\scshape i\kern-0.25em b}\kern-0.8em\TeX}}}

\copyrightyear{2024}
\acmYear{2024}
\setcopyright{rightsretained}
\acmConference[CHI EA '24]{Extended Abstracts of the CHI Conference on Human Factors in Computing Systems}{May 11--16, 2024}{Honolulu, HI, USA}
\acmBooktitle{Extended Abstracts of the CHI Conference on Human Factors in Computing Systems (CHI EA '24), May 11--16, 2024, Honolulu, HI, USA}
\acmDOI{10.1145/3613905.3650875}
\acmISBN{979-8-4007-0331-7/24/05}

\usepackage{todonotes}
\let\xtodo\todo
\renewcommand{\todo}[1]{\xtodo[inline,color=orange!75]{#1}}

\usepackage{subcaption}
\usepackage{xcolor}
\usepackage{makecell}
\begin{document}

\title[The Illusion of Performance]{The Illusion of Performance: The Effect of Phantom Display Refresh Rates on User Expectations and Reaction Times}


% \affiliation{%

% \affiliation{%
\author{Esther Bosch}
\orcid{0000-0002-6525-2650}
\affiliation{%
  \institution{German Aerospace Center, Institute for Transportation Systems}
  \city{Braunschweig}
  \country{Germany}
}
\email{esther.bosch@dlr.de}

\author{Robin Welsch}
\orcid{0000-0002-7255-7890}
\affiliation{%
  \institution{Aalto University}
  \city{Espoo}
  \country{Finland}}
\email{robin.welsch@aalto.fi}

\author{Tamim Ayach}
\orcid{0009-0007-7255-7890}
\affiliation{%
  \institution{HU Berlin}
  \city{Berlin}
  \country{Germany}}
\email{mohammed.tamim.ayach@student.hu-berlin.de}

\author{Christopher Katins}
\orcid{0000-0001-6257-7057}
\affiliation{%
  \institution{HU Berlin}
  \city{Berlin}
  \country{Germany}}
\email{christopher.katins@hu-berlin.de}

\author{Thomas Kosch}
\orcid{0000-0001-6300-9035}
\affiliation{%
  \institution{HU Berlin}
  \city{Berlin}
  \country{Germany}}
\email{thomas.kosch@hu-berlin.de}

\renewcommand{\shortauthors}{Bosch et al.}

\begin{abstract}
%   \todo{Answer each question with one sentence}
% \todo{What is the problem?}
% \todo{What is the solution?}
% \todo{what is the approach?}
% \todo{what is the result?}
% \todo{what is the conclusion in the bigger picture?}

User expectations impact the evaluation of new interactive systems. Increased expectations may enhance the perceived effectiveness of interfaces in user studies, similar to a placebo effect observed in medical studies. To showcase the placebo effect, we conducted a user study with 18 participants who performed a target selection reaction time test with two different display refresh rates. Participants saw a stated screen refresh rate before every condition, which corresponded to the true refresh rate only in half of the conditions and was lower or higher in the other half. Results revealed successful priming, as participants believed in superior or inferior performance based on the narrative despite using the opposite refresh rate. Post-experiment questionnaires confirmed participants still held onto the initial narrative. Interestingly, the objective performance remained unchanged between both refresh rates. We discuss how study narratives influence subjective measures and suggest strategies to mitigate placebo effects in user-centered study designs.

% User expectations play an important role when evaluating novel interactive systems. Increased user expectations can lead to an improved assessment of interfaces in user studies although the novel interface does not provide a real benefit, similar to a placebo effect known from medicine. We showcase this circumstance through a user study (N=18) where participants conducted a reaction time test with different refresh rates explained to the participants before each condition. Unbeknownst to the participant, we employed narrated refresh rates that were merely stated by the experimenter while participants were playing with either a slower or faster refresh rate. Our results show that the participants were successfully primed prior the experiment, believing in a superior or inferior performance although they were using the opposite refresh rate. Post questionnaires revealed that participants were still believing the narrative. Interestingly, no changes in the objective performance were found. We discuss how study narratives can manipulate subjective measures and conclude with recommendations to control for placebo effects in user-centered study designs.
\end{abstract}




\begin{teaserfigure}
  \includegraphics[width=\textwidth]{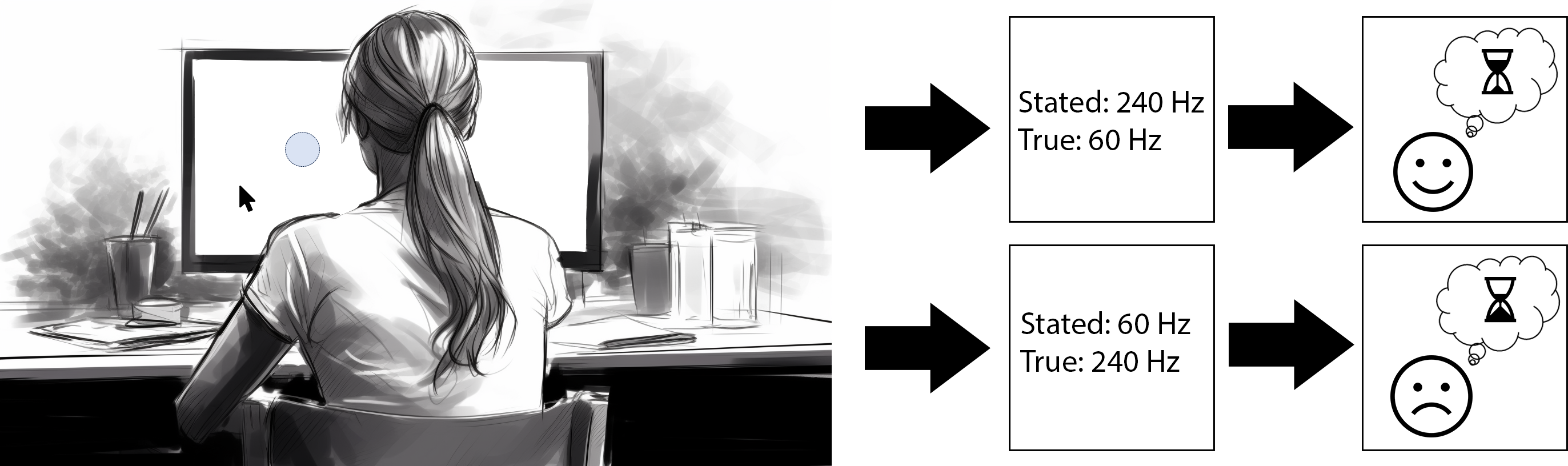}
  \caption{Users conducted a target selection reaction time task with different display refresh rates. Users were instructed beforehand about the refresh rate. However users received instructions about using a superior or inferior refresh rate while using the opposite or narrated refresh rate. Users who were allegedly using high refresh rates expected a higher performance. In contrast, users who obtained a low refresh rate expected a degraded performance even when using a high refresh rate.}
  \Description{In this work, users conducted a target selection reaction time task with different display refresh rates. Users were instructed beforehand about the refresh rate. However, using a full factorial study design with four conditions, users received instructions about using a superior or inferior refresh rate, although the users were using the opposite refresh rate. Users who were allegedly using high refresh rates expected a higher performance. In contrast, users who obtained a low refresh rate expected a degraded performance even when using a high refresh rate.}
  \label{fig:teaser}
\end{teaserfigure}


\maketitle

\section{Introduction \& Background}
A display's refresh rate determines how often a computer can show a new image per second. Nowadays, manufacturers produce consumer displays with refresh rates ranging between 60\,Hertz and 500\,Hertz (Hz). Higher refresh rates let content appear earlier on a computer display, reducing reaction times between users and the displayed content. Hence, computer displays with high refresh rates reduce the latency between user input and the viewed content, a critical gaming aspect. In this context, higher refresh rates reduce the overall reaction time~\cite{https://doi.org/10.1002/jsid.1198,  murakami2021study}. Yet, the refresh rate of a computer display is latent. For comparison, Halbhuber et al.~\cite{10.1145/3568444.3568448} showed that phantom network latencies affect players' subjective expectations and objective performance. Inspired by this work, this paper investigates if and how narrated but not placebic refresh rates influence user expectations and performance.

Interfaces in Human-Computer Interaction (HCI) tend to conceal the complexity of intelligent interface behavior. However, this can change users' expectations towards systems, potentially leading to biased study results~\cite{kosch2022placebo} or facilitating increased risk-taking when users believe that technologies improve their capabilities~\cite{villa2023the}. Parallel to medicine, research indicates that an individual's confidence in the efficacy of a treatment is enough to bring about enhancements in their physical or mental well-being, even when the treatment lacks genuine effectiveness. This positive change results from the patient anticipating the treatment will positively impact their health. This occurrence, where a person experiences benefits from an inactive intervention, is recognized as the placebo effect~\cite{MARGO199931}. A prerequisite is the expectation and user's belief in the efficacy of the treatment. A placebo, as evidenced by studies~\cite{lasagna1954study, beecher1955powerful, kaptchuk1998powerful}, can alleviate pain or contribute to the treatment of illnesses, providing an effective medical intervention despite lacking a specific mechanism for a particular ailment. The effectiveness of the placebo hinges on the patient's anticipation of its efficacy~\cite{Boot2013Psychplac}, resulting in positive post-treatment evaluations~\cite{stewart2004placebo, montgomery1996mechanisms}, and sometimes even objective physiological changes~\cite{enck2013placebo, enck2019does, schedlowski2015neuro}.
Consequently, the placebo effect complicates the assessment of new medical treatments, regardless of their actual utility. To address this challenge, medical studies incorporate placebos as controls during the evaluation of novel treatments. The effectiveness of a treatment is only acknowledged if the benefits surpass the improvements observed in participants treated with a placebo control condition. This placebo control approach is commonly applied in various scientific fields assessing human responses, such as psychological treatment~\cite{Boot2013Psychplac}, sports science~\cite{ojanen1994can}, and visualization research~\cite{correll2020actually}. However, this methodology is not standard when evaluating the efficacy of new interfaces in HCI.

HCI research has demonstrated that placebos can enhance usability and user experience without a functional system. Thus, controlling for user expectations in user studies gains importance in the HCI community~\cite{villa2024evaluating}. In gaming, fictitious power-up elements that do not impact gameplay and deceptive descriptions of AI adaptation have been found to boost self-reported game immersion~\cite{Denisova2019, denisova2015placebo}. Social media, offering control settings for prioritizing items in a news feed, can lead to higher subjective ratings of user satisfaction, even when these settings have no actual influence~\cite{vaccaro2018illusion}. Studies by \citet{kosch2022placebo} revealed that a non-functional supportive interface can induce a placebo effect related to perceived performance gains and reduced workload measures. In the context of participant beliefs about AI, \citet{pataranutaporn2023influencing} conducted a study where individuals interacting with a mental health AI chatbot were informed about different AI characteristics. Despite all participants engaging with the same AI model, those told that the AI was benevolent reported significantly higher levels of trustworthiness, empathy, and effectiveness in providing mental health advice than those primed to believe it was neutral or manipulative. Examining user expectations in human-AI interaction, \citet{kloft2023ai} found that increased expectations, irrespective of the actual presence of a supportive interface, improve performance due to placebo effects.
Interestingly, negative AI descriptions do not bias performance expectations. This dynamic can adversely affect human behavior when utilizing technologies believed to be enhancing. In addition, \citet{vicente2023inherited} showed that placebo effects manifest even after the interaction. \citet{villa2023the} explored the placebo effect within the context of human augmentation technologies, demonstrating a sustained belief of improvement after using a sham augmentation system and an increased willingness to take risks associated with heightened expectancy.

\begin{table}
    \centering
    \begin{tabular}{c c c}
        \toprule
          & \textbf{Stated 60 FPS} & \textbf{Stated 240 FPS} \\
         \midrule
         \textbf{True 60 FPS} & \makecell[c]{\textcolor{red}{Low (H1, H3)} \\ \textcolor{blue}{Low (H5)}} & \makecell[c]{\textcolor{red}{High (H1, H3)} \\ \textcolor{blue}{Low (H5)}}\\
         \\
         \textbf{True 240 FPS} & \makecell[c]{\textcolor{red}{Low (H1, H3)} \\ \textcolor{blue}{High (H5)}} & \makecell[c]{\textcolor{red}{High (H1, H3)} \\ \textcolor{blue}{High (H5)}} \\
         \bottomrule
    \end{tabular}
    \caption{Expected subjective assessments before and after the target selection reaction time task according to our hypotheses. We conducted a full factorial experimental design. \textbf{Red:} Subjective assessment before and after interaction. \textbf{Blue:} Objective reaction time.}
    \label{tab:hypotheses}
\end{table}
\begin{figure*}
  \begin{subfigure}{0.48\textwidth}
      \centering
      \includegraphics[height=0.21\textheight]{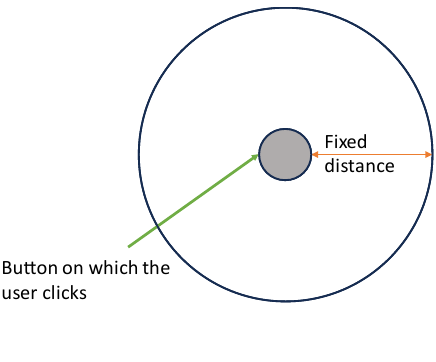}
      \subcaption{}
      \label{fig:task_left}
      \Description{}
  \end{subfigure}
  \hspace{\fill}
  \begin{subfigure}{0.48\textwidth}
      \centering
      \includegraphics[height=0.21\textheight]{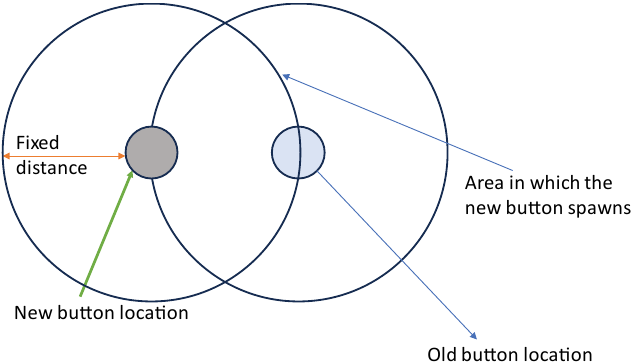}
      \subcaption{}
      \label{fig:task_right}
      \Description{}
  \end{subfigure}
  \caption{Description of the target selection reaction time task. \textbf{(a):} A button appears on which the participants must click as fast and accurately as possible. \textbf{(b):} A new button appears on the screen, and the old button disappears.}
  \label{fig:task}
\end{figure*}
Previous work showed that placebo effects exist when manipulating network latencies, showing positive effects when participants believe in playing a computer game with a low network latency~\cite{10.1145/3568444.3568448}. Inspired by this research, we investigate the users' performance under placebic refresh rates compared to their actual refresh rates (see \autoref{fig:teaser}). Participants were assigned to undergo a target selection reaction time task using a computer display that displayed refresh rates at 60\,Hz or 240\,Hz. Throughout four rounds, participants completed a reaction time task while being primed with a prior current refresh rate in use. However, the participants used either the true described refresh rate or a placebic stated refresh rate throughout the reaction time task. The results show that participants expected improved reaction time before experimenting using an allegedly high refresh rate. Interestingly, the participants still believed in their improved reaction time when using a low refresh rate in reality while using the placebo improvement narrative. However, no objective improvement in reaction times was found. Our research shows that the narrative of using technologies that improve human capabilities influences the assessment of users during studies, thus having wide implications on how the HCI community must deal with the assessment of technologies in the future.

\section{Methodology}
We conducted a within-subjects study design in a reaction time task to investigate the following hypotheses (see~\autoref{tab:hypotheses}) concerning different Frames per Second (FPS) for display refresh rates.
\begin{itemize}
    \item[\textbf{H1:}] Participants who will play with a true FPS of 60 Hz expect to perform better when told they will play with 240 Hz before playing the reaction time task.
    \item[\textbf{H2:}] Participants who will play with a true FPS of 240 Hz expect to perform worse when told they will play with 60 Hz before playing the reaction time task.
    \item[\textbf{H3:}] After playing with true 60 FPS, participants rated their performance better when told to play with 240 Hz.
    \item[\textbf{H4:}] After playing with true 240 FPS, participants rated their performance worse when told to play with 60 Hz.
    \item[\textbf{H5:}] Players' reaction times differ between 60 and 240 true frame rate.
\end{itemize}

\subsection{Task \& Experimental Setup}
We describe the target selection reaction time task in the following. The reaction time task is a target selection task. For the reaction time task, we designed a circular button that, upon clicking, relocates to a new position on the screen. To ensure the new location was not entirely randomized and comparable between the conditions and the participants, we implemented criteria dictating that the button's new position should be within a specific distance from its previous location (see \autoref{fig:task}). We opted for four different distances (i.e., 50px, 100px, 150px, 200px), randomly presenting the button at these distances in each round, with each distance occurring precisely five times. After clicking the displayed button, the next button appeared without delay in the defined vicinity of the previous button. Participants were required to complete the task over four rounds, each involving 20 button clicks. The participants conducted 80 trials throughout the study. Participants were primed with four narratives of a refresh rate overall before playing each round (see~\autoref{tab:hypotheses}). The button's size remained consistent, with 100px across all rounds. The task completion time was recorded for each click. The participants were instructed to click the button as fast and as accurately as possible. We used a Lenovo Y25-30 display that can run 60\,Hz and 240\,Hz using a computer that can render frames with the intended refresh rate for each condition. We verified the refresh rate correctly before starting each condition. We used a Razer Viper Ultimate with a polling rate of 1000\,Hz. We set the mouse to 200 dpi for each participant.

\subsection{Independent Variables}
Our experimental design consists of two independent variables. We define the \textit{stated refresh rate} as the refresh rate manipulated by the experimenter through a narrative. However, the real displayed refresh rate was a different one, i.e., 60\,Hz when the narrative was 240\,Hz and 240\,Hz when the stated refresh rate was 60\,Hz. In other words, the placebic refresh rate is not the actual refresh rate but a refresh rate that acts as a placebo. Similarly, we define the \textit{true refresh rate} with the rates 60\,Hz and 240\,Hz. Conditions with the same \textit{stated refresh rate} and \textit{true refresh rate} act as a baseline condition.

% words to explain: placebic refresh rate, post-game evaluation, actual refresh rate

\subsection{Dependent Variables}
We measure the reaction times between the appearance of the target item and the time the participant takes to click on the button. We calculate the expected performance rating using a five-point Likert scale (see~\autoref{tab:likert}) before and after playing each condition.

\begin{table*}[]
    \centering
    \begin{tabular}{l l}
        \toprule
         \textbf{ID} & \textbf{Question} \\
         \midrule
         Q1 &  I am using the 60\,Hz refresh rate, I think I will perform faster than the 240\,Hz refresh rate.\\
         Q2 & I am using the 240\,Hz refresh rate, I think I will perform faster than the 60\,Hz refresh rate. \\
         Q3 & I used the 60\,Hz refresh rate, I think I performed faster than the 240\,Hz refresh rate. \\
         Q4 & I used the 240\,Hz refresh rate, I think I performed faster than the 60\,Hz refresh rate.\\
         \bottomrule
    \end{tabular}
    \caption{Likert items used to assess the user expectations before (Q1 \& Q2) and after (Q3 \& Q4) each reaction time task trial.}
    \label{tab:likert}
\end{table*}

\subsection{Procedure}
This study employed a full factorial within-subjects design, conducting individual tests for each participant in four conditions. Participants were seated with a computer on the table and provided with an iPad equipped with an Apple pencil. An iPad presented an informed consent form in PDF format for participants to read and sign. Following this, participants were instructed to conduct one test trial of the reaction time task, which appeared on the entire screen and began with text outlining the task, algorithm, and a brief description of refresh rates. This trial aimed to acquaint participants with the task and allow them to request clarification from the instructor. Subsequently, participants were introduced to two question forms on the iPad. The first form featured Likert scale questions (see~\autoref{tab:likert}) about performance expectations for each refresh rate. Participants completed questions Q1 and Q2 before the respective condition and Q3 and Q4 after conducting the reaction time task. This ensures that the participants believe the narrative before the conditions, thus manipulating their expectations before each condition. While participants were informed of the refresh rate before each condition, they were unaware if they were playing with the stated or true refresh rate. Participants were then directed to select the refresh rate announced by the instructor for the reaction time test of that condition. After selection, the screen briefly turned black to simulate the system adjusting to the refresh rate. A 'click' button appeared, initiating the reaction time test. Afterward, participants completed a second question form on the iPad, evaluating their expected performance after the condition. This process was repeated three more times, resulting in four repetitions. The conditions were counterbalanced. Upon completion of the conditions, participants provided demographic information and details about their prior experience with refresh rates. We informed the participants about the experiment's deceptions and allowed them to withdraw their data from the experiment. The institute's review board gave ethical approval.

\subsection{Participants}
Overall, 18 participants were recruited for the study (nine female, eight male, and one preferred not to disclose their gender). The average age of the participants was 22.83 years (SD = 2.90). Eight participants were aware of the functionality of display refresh rates. Five participants heard about refresh rates but have yet to learn about the exact functionality. Five participants never heard of refresh rates before. When asked about their own used refresh rates, ten participants stated that they use a 60Hz refresh rate display, and three said that they use a 120\,Hz refresh rate display. Two participants stated that they use a 144\,Hz display. Two participants stated that they use a 165\,Hz monitor. Only one participant said they usually use a 240\,Hz refresh rate monitor.

\subsection{Data Analysis \& Results}

For H1 to H4, we tested the difference in performance assessment between the stated 60\,Hz and the stated 240\,Hz by the Wilcoxon signed-rank test. We corrected the alpha levels using Bonferroni correction for multiple tests. We calculated the effect size using the signed-r statistic. For H5, we calculated a regression that tested whether different stated and true refresh rate levels predict reaction times.

\subsection{User Expectations}

% (Significant?) effect between 60 and 240 Hz narrative, regardless of the acutal used refresh rate. We did 
% a regression

\begin{figure*}
  \includegraphics[width=\textwidth]{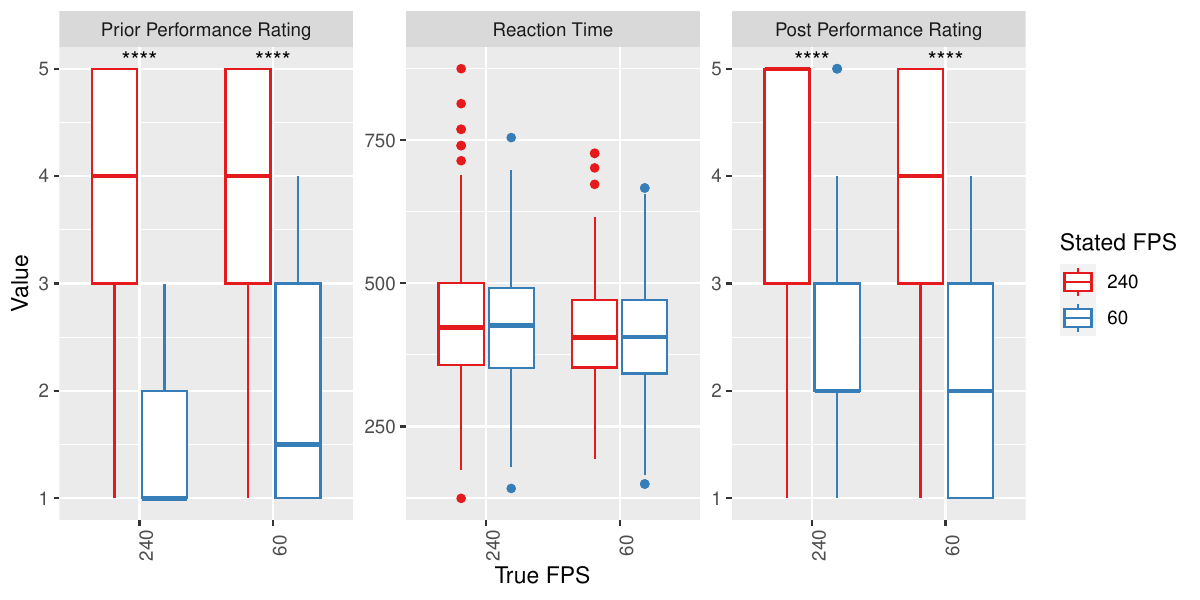}
  \caption{Aggregated user expectations before and post-interaction. Our narrative manipulated the user expectations towards the used refresh rates. Allegedly used high refresh rates elicited high user expectations before and after interaction, even when lower refresh rates were used. In contrast, a narrative of low refresh rates lead to lower performance expectations even when high refresh rates were used. No significant difference was found for objective reaction time measures.}
  \label{fig:box}
\end{figure*}

\autoref{fig:box} displays expected and evaluated performances by true and stated refresh rates. Participants who played with 60 Hz expected to perform significantly better when told they would play with 240\,Hz than with 60\,Hz, $Z = 32.5$, $p < .001$, $r = 0.88$. When participants played with 240\,Hz they expected to perform significantly better when they were told that they played with 240\,Hz than with\,60 Hz, $Z = 3.00$, $p < .001$, $r = 0.99$. 

After playing with 60\,Hz, participants rated their performance significantly better when they had been told to play with 240\,Hz rather than 60\,Hz, $Z = 42.50$, $p < .001$, $r = 0.86$. After playing with 240\ Hz, participants rated their performance significantly better when they were told to play with 240\,Hz than 60\,Hz, $Z = 16.50$, $p < .001$, $r = 0.92$.

\subsection{Reaction Times}
\autoref{fig:box} displays a descriptive plot of reaction times in different true and placebic refresh rates. However, placebic and true refresh rates did not significantly affect reaction times, $F(4,157) = 0.82$, $p = .51$.

\section{Discussion}
We conducted a study in which participants were presented with various narratives about display refresh rates while performing a reaction time task. Without the participants' knowledge, the refresh rates they were told differed from those used in reality. Although there were noteworthy variations in participants' anticipated ratings before and after the task, there was no significant distinction in reaction times between the stated and true refresh rates. In addition, objective reaction times did not differ between 60\,Hz and 240\,Hz true refresh rates. Although placebo effects can improve the user experience without providing a functional system, it is still being determined which factors study narratives must include to elicit placebo effects and for how long they are functional. This has severe implications in safety-critical scenarios, where users rely on improvements through interactive systems that may not contribute to safety. We discuss the impact of our results and the need to control for placebo effects in the following.

\subsection{User Expectations Manipulate Self-Assessments Prior Interaction}
Our results indicate that participants anticipated performance in line with the stated refresh rate before engaging in the reaction time test. Participants who were informed that they would be using a refresh rate of 240\,Hz rated their performance higher than when using a refresh rate of 60\,Hz. Conversely, participants rated their performance lower when informed they would be using 60\,Hz instead of 240\,Hz. This pattern was consistent for both the stated and true refresh rates. The statistical significance of the ratings before interaction supports the confirmation of \textbf{H1} and \textbf{H2}. Our findings align with previous research examining the impact of artificial intelligence priming on anticipated user performance~\cite{kloft2023ai, kosch2022placebo} and human augmentation~\cite{villa2023the}. These studies suggest that when users are unaware of a system's improved functionality that is not easily perceptible, their expectations tend to increase. Consequently, these heightened expectations may bias subjective scores when users are unaware of an improvement. Therefore, assessing subjective user performance ratings before interaction indicates user expectations and should be approached cautiously, considering the potential for placebo effects.

\subsection{User Expectations Manipulate Self-Assessments Post Interaction}
Moreover, we assessed the performance rating following the interaction with the reaction task game. Similar to the prior subjective performance ratings, the subjective performance scores exhibited significant differences even when participants engaged with a stated refresh rate that differed from the true refresh rate. This suggests that participants continued to hold onto the performance improvement or degradation post-interaction narrative. As a result, we substantiate the validity of \textbf{H3} and \textbf{H4}. This aligns with prior research demonstrating that voice assistants, whether portrayed as benevolent or malevolent when neutral, contribute to biased participant perceptions after interaction~\cite{pataranutaporn2023influencing}, indicating sustained effects post-interaction~\cite{vicente2023inherited}. 

\subsection{Objective User Performance}
We assessed the performance in the reaction task for each condition, which was measured in time per click trial. Our statistical analysis did not reveal a significant effect between the conditions. Similar findings were reported in other studies, where no direct differences in objective performance were observed~\cite{kosch2022placebo, DENISOVA201956}. Thus, we were unable to support \textbf{H5}. Interestingly, in contrast, another study identified significant differences in gaming performance when presenting phantom latencies as network latency in shooter games~\cite{10.1145/3568444.3568448} or when using sham-AIs combined with nocebo conditions~\cite{kloft2023ai}. Potential placebo effects in objective game performance may be task-dependent and influenced by specific task-related designs or placebo display configurations. For example, changing the task design to include moving elements while using different refresh rates may change the objective task performance~\cite{10049694}. Future research will investigate the design parameters of placebo indicators and their impact on task performance.

\subsection{Controlling for Placebo Effects in User-Centered Studies}
Our findings highlight the significance of evaluating the user's subjective ratings before and after a study condition to determine the susceptibility of a system with latent functionalities to placebo effects. Examining the performance expectations before and after interaction and comparing scores between stated and actual refresh rates effectively indicate the system's vulnerability to placebo effects. This approach aligns with the recommendations from prior research~\cite{Boot2013Psychplac}, emphasizing the importance of assessing user expectations before and after each experimental condition to mitigate biases stemming from user expectations in each condition. Thus, studies should include a placebo condition to test from biased expectations based on study narratives.

\section{Conclusion}
Display manufacturers provide users with various display options featuring different refresh rates. Users often opt for displays with higher refresh rates based on their intended use, expecting enhanced performance in their tasks. In our study, we showed the participants' expectations to be manipulated according to the stated refresh rate, even when dealing with placebic-stated refresh rates rather than the true refresh rate employed in reality. While our experiment successfully manipulated the participant's expectations when playing with different refresh rates, the reaction time did not show a significant effect. Nevertheless, our research demonstrates an approach for investigating narratives that claim to enhance or reduce perceived system functionalities, affecting user performance ratings. The manipulation of user expectations and the active control of other subjective measures in future HCI research pose a challenge for future research.

\bibliographystyle{ACM-Reference-Format}
\bibliography{main}

\end{document}